# Superconducting properties and Hall Effect of epitaxial NbN thin films

S P Chockalingam[a], Madhavi Chand[a], John Jesudasan[a], Vikram Tripathi[b] and Pratap Raychaudhuri[a*]

[a)]Department of Condensed Matter Physics and Materials Science, Tata Institute of Fundamental Research, Homi Bhabha Rd., Colaba, Mumbai 400005, India.
[b)]Department of Theoretical Physics, Tata Institute of Fundamental Research, Homi Bhabha Rd., Colaba, Mumbai 400005, India.

***Abstract:*** We have measured the magnetotransport properties and Hall effect of a series of epitaxial NbN films grown on (100) oriented single crystalline MgO substrate under different conditions using reactive magnetron sputtering. Hall effect measurements reveal that the carrier density in NbN thin films is sensitive to the growth condition. The carrier density increases by a factor of 3 between the film with highest normal state resistivity ($\rho_n \sim 3.83 \mu\Omega$-m) and lowest transition temperature ($T_c \sim 9.99K$) and the film with lowest normal state resistivity ($\rho_n \sim 0.94 \mu\Omega$-m) and highest transition temperature ($T_c \sim 16.11K$) while the mobility of carriers does not change significantly. Our results show that the $T_c$ of NbN is governed primarily by the carrier density rather than disorder scattering. By varying the carrier concentration during growth we can vary the effective disorder ($k_F l$) from the moderately clean limit to the dirty limit which makes this system ideal to study the interplay of carrier density and disorder on the superconducting properties of an s-wave superconductor.

---
[*] Electronic mail: pratap@tifr.res.in

## I. Introduction

Superconductivity in NbN thin films has been widely studied in the past decades motivated by diverse applications such as Josephson junctions[1], superconducting hot electron bolometers[2] and superconducting single photon detectors[3]. Within conventional superconductors NbN has high critical temperature ($T_c \sim 16K$) combined with a short coherence length ($\xi < 5nm$) and large penetration depth ($\lambda \sim 200nm$), which allows fabrication of few nanometer thick superconducting thin films with moderately high $T_c$. In addition, NbN grows as epitaxial thin films on lattice matched substrates using popular thin film growth techniques such as reactive magnetron sputtering[4,5,6,7,8,9] and pulsed laser deposition[10,11]. These films have good mechanical strength, are chemically stable in ambient atmosphere and can be recycled from cryogenic temperatures to room temperature without any detectable degradation in their superconducting properties. Also, the existence of several other nitrides with widely varying properties such as the ferromagnet GdN and the insulator AlN, opens the possibility of exploring fundamental issues such as superconductor/ferromagnet proximity effect[12] and the role of interlayer coupling[13] on the superconducting properties of superconductor/insulator/superconductor multilayers.

Although there are several properties which are favorable for applications, the $T_c$ of NbN thin films is very sensitive to the growth conditions. Thin films with $T_c \sim 16K$ can only be synthesized under carefully optimized deposition parameters. Away from the optimal growth condition, the $T_c$ of the film decreases rapidly[4,5,6,7,8,9,10,11] to 10K or below. It has been shown that there is a direct correlation between the normal state resistivity ($\rho_n$) and the $T_c$ [ref. 5]: Films with higher $\rho_n$ have lower $T_c$. Such a direct correlation between $\rho_n$ and $T_c$ for an s-wave superconductor is not straightforward. While it is natural to attribute the increase in $\rho_n$ to the increase in disorder in films grown away from the optimal condition, it was shown by Anderson[14] that scattering from non-magnetic disorder is not expected to significantly affect the $T_c$ in an s-wave superconductor, as long as the system remains a metal[15,16,17]. On the other hand $\rho_n$ will also change if the carrier density is sensitive to the growth condition. In this case, the superconducting $T_c$ which is sensitive to the density of states ($N(0)$) at Fermi level, is also expected to change. This possibility, which can be resolved through a determination of the carrier density through Hall effect in thin films with different $T_c$, has not been experimentally explored.

In this paper, we examine the role of carrier density and disorder on the superconducting properties of a series of epitaxial NbN films with $T_c$ varying from 9.99K to 16.11K. From measurements of the normal state and superconducting properties we extract the key parameters, such as the carrier density ($n$), the Fermi wave vector ($k_F$) the mean free path ($l$) and coherence length ($\xi_{GL}$) for all the films. The central observation of this paper is that the carrier density ($n$) extracted from Hall effect, varies monotonically from $n\sim6.47\times10^{28}$ electrons/m$^3$ in the film with lowest $T_c$, to $n\sim1.98\times10^{29}$ electrons/m$^3$ in the film with the highest $T_c$. Analysis of the normal state and superconducting properties reveal that the $T_c$ of NbN films is governed primarily by the carrier density rather than disorder scattering. However, the disorder characterized by the Ioffe-Regel parameter,[18] $k_Fl$, varies from moderately clean limit ($k_Fl\sim7.15$) to the dirty limit ($k_Fl\sim2.56$) making this system ideal to look at the interplay between carrier density and disorder in an s-wave superconductor.

## II. Experimental Details

Thin films of NbN were synthesized through reactive dc magnetron sputtering by sputtering an Nb target in Ar-N$_2$ gas mixture. The substrate temperature and ambient pressure during growth for all the films were fixed at $600^0$C and 5 mTorr respectively. Films with different $T_c$ were synthesized using two different protocols. The first set of films (Series 1) was synthesized by keeping the $N_2$/Ar ratio fixed at, 20% $N_2$-80% Ar, while the sputtering power on the Nb target was varied from 40W to 250W. These films are named as 1-NbN-*x*, where "*x*" refers to the sputtering power, e.g. 40 for 40W. The second set of films (Series 2) were synthesized by keeping the sputtering power at 80W and varying the $N_2$/Ar ratio from 10% $N_2$-90% Ar to 30% $N_2$-70% Ar. These films are named as 2-NbN-*y* where "*y*" refers to the $N_2$ partial pressure, e.g. 30 for 30% $N_2$-70% Ar mixture. The thickness of all the films was larger than 50nm to avoid the effect of epitaxial strain significantly affecting the structural and superconducting properties. The films were characterized through X-ray diffraction $\theta-2\theta$ scans using Cu K$_\alpha$ source. The epitaxial nature of the films was confirmed by performing $\phi$-scans around the [111] peak on a 4 circle X-ray goniometer. The temperature dependence of resistivity ($\rho(T)$) and $T_c$ was measured through conventional ac 4-probe technique in a home built cryostat up to a maximum field of 5.8T. The upper critical field was determined from $\rho(T)-T$ scans in different applied magnetic fields (magnetic field perpendicular to film plane) and determining the $T_c$ at various fields. To accurately determine $\rho(T)$, the electrical resistance was measured on a 1mm by 10mm stripline deposited using a shadow mask. The thickness of the film was measured using a Ambios X P 2 Stylus Profilometer at various positions on the stripline and the average value was taken as the film thickness. The thickness measured on different parts had a variation of ~15%. However, since this figure is close to the resolution limit of our thickness measurements, this possibly does not reflect the intrinsic level of non-uniformity in the thickness. This is likely to be the biggest source of error in the calculation of resistivity and Hall coefficient. The Hall measurement was carried out using 4-probe ac technique. The Hall voltage was deduced from reversed field sweeps from +5.8T to -5.8T after subtracting the resistive contribution.

## III. Results

Figure 1(a) and (b) show the representative X-ray diffraction $\theta-2\theta$ scans for the films in both the series, plotted in logarithmic scale for the intensity. All the films other than 1-NbN-250 and 2-NbN-15 show a peak corresponding to (200) peak of NbN which crystallizes in the *fcc* structure. The end member of Series 1 (1-NbN-250), namely, the film grown at the highest sputtering power shows a prominent diffraction peak at $38.63^0$ corresponding to the most intense peak (101) of $Nb_2N$ with no peak corresponding to NbN. This is not unexpected since the effect of increasing the sputtering rate of Nb at a fixed $N_2$ partial pressure, would increase the Nb/N ratio. Therefore the stability of the NbN phase should be limited by a maximum power consistent with our observation. 2-NbN-15 on the other hand shows only a very diffuse peak corresponding to NbN. For this film the superconducting transition is broad ($\Delta T_c \sim 2.5K$) and the resistance below the transition temperature decreases to 0.1% of its value in the normal state, but does not go to zero within our measurement resolution. We therefore do not consider this film in the rest of our analysis. For NbN films other than these two, we observe a very small amount of $Nb_2N$ impurity peak. We estimate the $Nb_2N$ impurity phase to be 0.2-0.4 volume % of the NbN phase based on the ratio of the area under the peaks. The lattice parameters (*a*) of the NbN films calculated from (200) peak is shown in figure 1(c). The epitaxy of all the films was confirmed from $\phi$–scan measurements on a 4-circle X-ray goiniometer.

Figure 2(a) shows the $\rho(T)$ vs. T for all the films. The normal state electrical resistivity ($\rho_n$) of the films measured at 17K varies from 0.94$\mu\Omega$-m for the least resistive film to 3.83 $\mu\Omega$-m for the most resistive film. The $T_c$ of the film is determined from the temperature at which the resistance falls to 10% of the normal state value. The inverse correlation between $T_c$ and $\rho_n$ is shown in figure 2(b). It is however instructive at this stage, to look separately at the $\rho_n$ and $T_c$ of films (shown in Fig. 2(c) and Fig. 2(d)) in Series 1 and Series 2. For Series 1, the NbN film with smallest $\rho_n$ and highest $T_c$ forms close to the boundary between $Nb_2N$ and NbN phase. In series 2, $T_c$ increases rapidly with increasing $N_2$ at low $N_2$ partial pressure followed by a gradual decrease at higher pressures. Our observation in Series 2 is consistent with the $N_2$ partial pressure dependence of $T_c$ reported by other groups[5,6].

Figure 3 shows the Hall resistivity ($\rho_{xy}$) as a function of magnetic field for all the films measured at 17K. The Hall coefficient, ($R_H = \dfrac{\rho_{xy}}{H}$) is determined from the slope of the $\rho_{xy}$-H curve. It is interesting to note that the carrier density ($n_e = \dfrac{1}{R_H e}$) extracted from $R_H$ for the film with $T_c \sim 16.11$K (1-NbN-200), $n_e = 1.97 \times 10^{29}$ electrons/$m^3$, is close to the theoretical estimate[19] of $2.39 \times 10^{29}$ electron/$m^3$. This agreement is remarkably good considering that the theoretical estimate was obtained by counting the number of electrons outside the closed shell of Nb under the assumption that the nitrogen atoms do not contribute. However, for films with lower $T_c$, $n_e$ decreases significantly reaching a value $n_e = 6.47 \times 10^{28}$ electrons/$m^3$ for the films with $T_c \sim 9.99$K (2-NbN-30).

Figure 4(a) shows the representative $\rho(T,H)$-T plots for the 1-NbN-200 films measured in different magnetic fields. The $T_c(H)$ at different magnetic field is determined from the temperature where the $\rho(T)$ drops to 10% of the normal state value. The upper critical field ($H_{c2}$) as a function of temperature is plotted in figure 4(b) by inverting the $T_c(H)$ data for different films.

The values of $a$, $T_c$, $\rho_n$, $n_e$ and $H_{c2}$(T/T$_c$=0.9) for all the films are listed in Table 1.

**Discussion**

Table 2 lists the Fermi wave vector ($k_F$), the Fermi velocity ($v_F$), the electronic mean free path ($l$), the density states at Fermi level ($N(0)$), the Ioffe-Regel parameter $k_F l$, the upper critical field at T=0 ($H_{c2}(0)$), and the Ginzburg Landau coherence length ($\xi_{GL}$) for all the films. $k_F$, $v_F$, $l$ and $N(0)$ are calculated from the free electron relations, $k_F = \left(3\pi^2 n\right)$, $v_F = \dfrac{\hbar k_F}{m}$ and $\rho = \dfrac{m}{ne^2\tau} = \dfrac{mv_F}{ne^2 l}$ and $N(0) = \dfrac{m}{\hbar^2\pi^2}\left(3\pi^2 n\right)^{1/3}$ where $m$ is the mass of the electron and $\tau$ is the relaxation time. To check the reliability of the parameters obtained from the free electron relations we compare $N(0)$ with the value obtained from electronic structure calculations as well as specific heat measurements. Using the augmented plane wave method Mattheiss[20] calculated the density of states for stoichiometric NbN to be 0.54 states/NbN-eV. From specific heat measurements on NbN$_{1-x}$ single crystals Geballe et al.[21] reported $N(0)\approx 0.5$ states/NbN-eV for the stoichiometric compound. For the film with $T_c$=16.1K, for which the $n$ is closest to value expected for the stoichiometric compound, the carrier density obtained from our measurements corresponds to $N(0)\approx 0.515$ states/NbN-eV which is in excellent agreement with these results. The consistency of $N(0)$ obtained using the free electron model may be traced back to very short mean free path of the NbN films (3.96Å for the film with $T_c\sim16.1$K). The presence of a large amount of disorder scattering is expected to smear out any fine structure in the electronic density of states leaving only a free electron like structure. This argument will hold even for the other films with lower $T_c$ where the mean free path is shorter. Since all our films are in the dirty limit, $l<<\xi_{GL}$, $H_{c2}(0)$ and $\xi_{GL}$ are estimated from the dirty limit relation[22],

$$H_{c2}(0) = 0.69 T_c \frac{dH_{c2}}{dT}\bigg|_{T=T_c} \text{ and } \xi_{GL} = \left[\Phi_0 / 2\pi H_{c2}(0)\right]^{1/2}. \tag{1}$$

Figure 5 (a) shows the dependence of $T_c$ on the carrier density $n$. $T_c$ increases monotonically with increasing $n$. It is interesting to note that we do not see any correlation between the $T_c$ and the lattice parameter of the film when all the films are taken into account[23]. Figure 5(b) shows the variation of conductivity at 17K, $\sigma_n (=1/\rho_n)$, as a function of $n$. The conductivity follows a linear trend with, $\sigma_n \propto n$, showing that the mobility of the carriers is not significantly different in different films[24]. The increase in $\rho_n$ in films with lower $T_c$ is therefore predominantly due to the decrease in carrier density rather than a change in scattering rate of the electrons. It therefore looks likely that the $T_c$ in different films is also primarily governed by the change in $n$ rather than disorder scattering.

To understand the dependence of $T_c$ on $n$ beyond this qualitative picture, we analyze our data within the framework of McMillan theory[25] for strong coupling superconductors[26]. Within this theory $T_c$ is given by,

$$T_c = \frac{\Theta}{1.45}\exp\left(-\frac{1.04(1+\lambda)}{\lambda - \mu^*(1+0.62\lambda)}\right), \tag{2a}$$

where $\lambda$ is the electron-phonon coupling constant, $\mu^*$ accounts for the electron-electron repulsive interaction and $\Theta$ is of the order of Debye temperature. Within McMillan theory $\lambda$ is proportional to the density of states at Fermi level ($N(0)$),

$$\lambda = N(0)\frac{\langle g^2 \rangle}{M\langle \omega^2 \rangle} = KN(0), \tag{2b}$$

where $K = \dfrac{\left\langle g^2 \right\rangle}{M \left\langle \omega^2 \right\rangle}$, $<g^2>$ is an average over the square of the electron-phonon interaction matrix elements and $<\omega^2>$ is the average of the square of phonon frequency and M is the mass of the ion. To fit our data to the McMillan equations we rewrite (2a) and (2b) as

$$\ln(T_c) = \ln\left(\frac{\Theta}{1.45}\right) - \frac{1.04(1 + KN(0))}{KN(0) - \mu^*(1 + 0.62KN(0))}, \qquad (3)$$

We fit the variation of ln($T_c$) with the *N(0)* using $\ln\left(\dfrac{\Theta}{1.45}\right)$ and $K$ as fit parameters. While making this fit we assume that $\Theta$ and K are same for all films. This assumption is justified since the lattice constants for different films do not vary by more than 0.34% across both the series and hence we do not expect the phonon spectrum and the electron-phonon interaction matrix elements to change significantly for different films. In transition metal compounds the value of $\mu^*$ is taken to be, $\mu^*$=0.13 [ref. 25,27]. Figure 6 shows the fit of our data with equation (3). Barring one point all the points fall very close to the fitted plot within the error bars. The extracted value, $\Theta \approx 174K$, is in the same ballpark as the reported values of Debye temperature[27,21,28,29] which vary in the range 250-320K[30]. The value of $\lambda$ (listed in Table 2) calculated from the best fit value of $K$ varies between $\lambda$=1.13-1.64 consistent with the strong coupling nature of NbN. These values are in qualitative agreement[31] with the value $\lambda$=1.45 obtained by Kihlstrom et al. from tunneling data on a film with $T_c$=14K. Despite this agreement there are some points of caution in the preceding analysis. First, it has been suggested by Maekawa et al.[32] that for large disorder could cause a decrease in *N(0)* from its free electron value due to the loss of effective screening. Though our samples with lower $T_c$ have $k_F l$ in the range where this effect could manifest itself we do not observe such an effect from our data either in the normal[33] or the superconducting state. Secondly, the value of $\mu^*$ has to be critically assessed for disordered superconductors. It was argued by Anderson et al.[17,34] that for large disorder, the loss of effective screening can result in an increase in $\mu^*$ by an amount of the order of $\mu^*/(k_F l)^2$. To explore this possibility we have also fitted the data assuming $\mu^*$=0.16. While the value of $\Theta$ remains the same, the value of $\lambda$ increases by about 10% which is well within the error that we expect from this fitting procedure.

We now focus our attention to the $H_{c2}(0)$ for all the films. For dirty superconductors $H_{c2}(0)$ is related to $T_c$ and $N(0)$ through the expression[22] (in SI units),

$$H_{c2}(0) = 0.69 T_c \frac{4 e k_B}{\pi} N(0) \rho_n.$$  (4)

Figure 7 shows the experimental values of $H_{c2}(0)$ along with the calculated values using equation (4) as a function of $\rho_n$. Despite qualitative similarities there are significant differences. The value of $H_{c2}(0)$ calculated using equation (4) is significantly smaller than the experimental value for all the films. In addition, the calculated value of $H_{c2}(0)$ increases with $\rho_n$ above 2$\mu\Omega$-m, whereas the experimental value actually decreases. One source of error in the calculated value could be the use of the free electron density of states for $N(0)$. This however seems unlikely since $N(0)$ is consistent with specific heat measurements. We would like to note that there could be several sources of error in the values of $H_{c2}(0)$ extracted from equation (1). First, since the superconducting transition becomes broad in the presence of magnetic field, the experimental values of $T_c(H)$, depends very much on the criterion used to determine $T_c(H)$. To explore this possibility we have also calculated $H_{c2}(0)$ by determining $T_c(H)$ from the value at which the resistivity is 50% of the normal state value (Figure 7). While this changes the experimental values of $H_{c2}(0)$ by about 20% for films with higher $\rho_n$ the overall trend does not change significantly. Secondly, equation (1) is not universally valid. Significant deviation from equation (1) has been observed in dirty systems, where the critical field has been observed to increase linearly[35] with decreasing temperature down to $0.2T_c$. Also spin-orbit interactions can cause significant enhancement of $H_{c2}(0)$ over the value[22] given by equation (4). However, we believe that a more direct measurement of the critical field down to low temperatures at high fields is necessary to reliably address this issue.

We can try to speculate the reason behind the large change in carrier density on the deposition conditions in NbN. Since these films are grown by changing the sputtering power or $N_2$ partial pressure it is expected that they differ in their Nb/N ratio. Previous studies on the dependence of electronic properties of NbN on stoichiometry were carried out on single crystals with nitrogen[21] vacancies. While we were unable to quantitatively determine the Nb/N ratio in these films using conventional techniques, nitrogen vacancies alone cannot explain our results. In Series 1, the film with highest $T_c\sim16.11K$ forms close to the phase boundary between the $Nb_2N$ and the NbN phase. Since this film has a carrier density which is close to the expected value for the stoichiometric compound, it seems likely that this film has stoichiometry close to the Nb:N ratio of 1:1. In this series $n$ (and $T_c$) decreases when we decrease the sputtering power. Since decreasing the sputtering rate of Nb decreases the $Nb:N_2$ ratio in the plasma, films grown at lower sputtering power should have lower Nb content. Therefore films with lower $n$ and $T_c$ are likely to have Nb vacancies. (Similar argument will also hold for Series 2 where increase in $N_2$ partial pressure above 20% results in a decrease in $n$ and $T_c$ in the range where the NbN crystallographic phase is stable.) Since the conduction electrons in this material come primarily from the electrons outside the close shell of Nb an increase in Nb vacancies in the films would cause a sharp decrease in $n$. However, to account for a three-fold decrease in $n$ we would need a very large degree of Nb vacancies which is possibly unacceptable from the structural point of view. Since N vacancies also reduce the $n$ and $N(0)$ in the sub-stoichiometric compound[36,37], it is possible that the drastic reduction of carrier density is caused by a combination of Nb and N vacancies[38]. In addition we cannot rule out the possibility of non-trivial modifications of the band structure caused by the presence of both Nb and N vacancies. This will require detailed calculations of the electronic structure beyond the rigid band picture, in the presence of Nb and N vacancies. Such calculations do not exist at present and is beyond the scope of this paper.

Finally, we would like to note that the ability to control the carrier density in NbN thin films provides a unique opportunity to look at the interplay of carrier density and disorder in an s-wave superconductor without any external doping. It is interesting to note the Ioffe-Regel parameter, $k_F l$, which is a measure of the mean free path in terms of the de-Broglie wavelength of the electron at Fermi level, varies from 2.56 to 7.14. Since $k_F l \sim 1$ corresponds to the Mott limit for metallic conductivity, the level of disorder varies from moderately clean to the dirty limit. This provides an opportunity to look at the role of dynamic fluctuations as a function of disorder, as well as to probe recent theoretical predictions[15] of the formations of pseudogap state where the weak disorder limit breaks down. At least one such study has already been attempted[39] and we believe that our work will provide a much better understanding to those data. Our own investigations in this direction are currently underway and will be reported elsewhere.

**Conclusion**

We have explored the interplay of carrier density and disorder in epitaxial NbN films grown using dc magnetron sputtering. Our results show that the $T_c$ in these films is primarily governed by the carrier density which increases by a factor of 3 as we go from the film with lowest $T_c$~9.99K to the films with highest $T_c$~16.11K. By fitting our data with the McMillan theory we obtain the value of the electron-phonon coupling parameter λ~1.07−1.55, consistent with the expected value for strong coupling superconductors.

***Acknowledgements:*** We would like to thank Profs. Nandini Trivedi, S. Ramakrishnan, K. L. Narasimhan, Pushan Ayyub and T. V. Ramakrishnan for useful discussions. We would like to thank Mahesh Gokhale, Abdul Kadir and Pappu Laskar for their help with measurement on the 4-circle goniometer and B. A. Chalke, Nilesh Kulkarni and Vivas Bagwe for technical help.

**Table 1.** Lattice parameter (*a*), Superconducting transition temperature (*T_c*) normal state resistivity ($\rho_n$), electron density (*n*) and *H_{c2}*(t =T/T_c=0.9) for all the films in Series 1 and Series 2. Note that 1-NbN-80 and 2-NbN-20 corresponds to the same film.

| Sample | *a* (Å) | *T_c* (K) | $\rho_n$(17K) ($\mu\Omega$-m) | *n* (electrons/$m^3$) | *H_{c2}* at T/T_c = 0.9 (T) |
|---|---|---|---|---|---|
| 1-NbN-200 | 4.4130 | 16.11 | 0.9459 | $1.97 \times 10^{29}$ | 2.16 |
| 1-NbN-150 | 4.4036 | 15.14 | 1.4787 | $1.61 \times 10^{29}$ | 2.59 |
| 1-NbN-100 | 4.4107 | 13.57 | 1.8914 | $9.28 \times 10^{28}$ | 2.54 |
| 1-NbN-80 | 4.4041 | 12.24 | 1.9401 | $8.72 \times 10^{28}$ | 2.22 |
| 1-NbN-40 | 4.3980 | 11.61 | 2.2575 | $8.00 \times 10^{28}$ | 1.77 |
| 2-NbN-15 | * | 10.73[q] | * | * | * |
| 2-NbN-25 | 4.4160 | 11.27 | 3.1203 | $7.60 \times 10^{28}$ | 1.59 |
| 2-NbN-30 | 4.4260 | 9.99 | 3.8311 | $6.46 \times 10^{28}$ | 1.59 |

[q] For this film the resistance decreases to 0.1% of its normal state value below the transition but does not go to zero. We list it here only to show the overall trend of *T_c* with $N_2$ partial pressure.

**Table 2.** The Fermi wave vector ($k_F$), Fermi velocity ($v_F$), electronic mean free path ($l$), the upper critical field ($H_{c2}(0)$) and the Ginzburg Landau coherence length ($\xi_{GL}$), density of states at Fermi level ($N(0)$), Ioffe-Regel parameter ($k_F l$) and electron phonon coupling strength ($\lambda$) for all the films.

| Sample | $k_F$ (m$^{-1}$) | $v_F$ (m/s) | $l$ (Å) | $H_{c2}(0)$ (T) | $\xi_{GL}$ (nm) | $N(0)$ (states/m$^3$-eV) | $k_F l$ | $\lambda$ |
|---|---|---|---|---|---|---|---|---|
| 1-NbN-200 | $1.80 \times 10^{10}$ | $2.09 \times 10^6$ | 3.96 | 14.80 | 4.72 | $2.38 \times 10^{28}$ | 7.15 | 1.64 |
| 1-NbN-150 | $1.68 \times 10^{10}$ | $1.95 \times 10^6$ | 2.91 | 17.82 | 4.30 | $2.23 \times 10^{28}$ | 4.90 | 1.53 |
| 1-NbN-100 | $1.40 \times 10^{10}$ | $1.62 \times 10^6$ | 3.28 | 17.58 | 4.33 | $1.86 \times 10^{28}$ | 4.60 | 1.28 |
| 1-NbN-80 | $1.37 \times 10^{10}$ | $1.59 \times 10^6$ | 3.34 | 16.65 | 4.45 | $1.82 \times 10^{28}$ | 4.58 | 1.25 |
| 1-NbN-40 | $1.33 \times 10^{10}$ | $1.54 \times 10^6$ | 3.04 | 15.12 | 4.67 | $1.77 \times 10^{28}$ | 4.05 | 1.21 |
| 2-NbN-25 | $1.31 \times 10^{10}$ | $1.52 \times 10^6$ | 2.28 | 14.79 | 4.72 | $1.74 \times 10^{28}$ | 2.98 | 1.19 |
| 2-NbN-30 | $1.24 \times 10^{10}$ | $1.44 \times 10^6$ | 2.07 | 13.08 | 5.02 | $1.65 \times 10^{28}$ | 2.56 | 1.13 |

**Figure captions:**

**Figure 1.** (color online) X-ray diffraction $\theta$-$2\theta$ traces of epitaxial NbN films plotted in logarithmic scale for intensity: (a) Series 1 and (b) Series 2. Successive plots are shifted upwards for clarity. All films other than 1-NbN-250W show the (200) NbN peaks. The peak marked * correspond to $Nb_2N$ impurity phase. (c) Lattice parameters as a function of sputtering power/$N_2$ partial pressure for NbN films in Series 1 and Series 2 respectively.

**Figure 2.** (color online) (a) Resistivity ($\rho$) as a function of temperature for all NbN films used in this study. (b) Normal state resistivity $\rho_n$ measured at 17K as a function of $T_c$. (c) $\rho_n$ (▲) and $T_c$ (■) as a function of sputtering power for NbN films in Series 1. (d) $\rho_n$ (▲) and $T_c$ (■) as a function of $N_2$ partial pressure for NbN films in Series 2; the sample 2-NbN-15 (marked as *) does not show zero resistance below the transition. The notional $T_c$ is shown in panel (d) to indicate the overall trend in $T_c$ with $N_2$ partial pressure.

**Figure 3.** (color online) Hall resistivity $\rho_{xy}$ as a function of H for all the films measured at 17K. The Hall coefficient ($R_H$) is obtained from the slope of $\rho_{xy}$-H curve by fitting a straight line.

**Figure 4.** (color online) (a) Resistivity ($\rho$) as a function of temperature for 1-NbN-200 film in different applied magnetic field. The successive magnetic field values are 0, 0.25T, 0.5T, 0.75T, 1T, 1.5T, 2T, 3T, 4T, 5T and 5.8T. (b) Upper critical field ($H_{c2}$) as a function of temperature (T) for all NbN films.

**Figure 5.** (color online) Variation of (a) $T_c$ and (b) $\sigma_n$ as a function of carrier density, $n$. The straight line in (b) shows the linear fit to the data with the relation $\sigma_n=Kn$, where K is a constant.

**Figure 6.** (color online) $\ln(T_c)$ as a function of the density of states at Fermi level, N(0). The line shows a fit to the data with the McMillan theory for strong coupling superconductors.

**Figure 7.** (color online) Measured (▲,●) and calculated (■) upper critical field, $(H_{c2}(0))$ as a function of $\rho_n$. ▲ and ● corresponds to two different criteria used to determine $T_c(H)$.

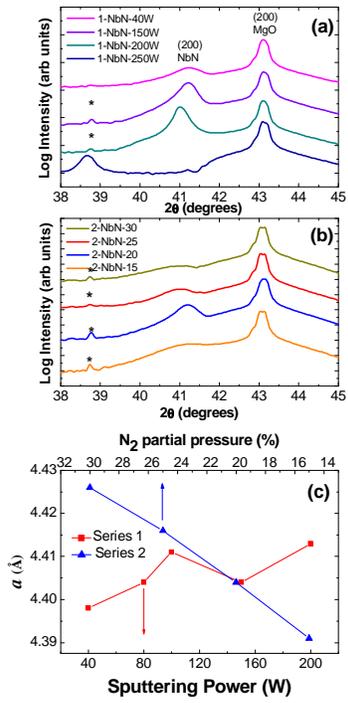

Figure 1

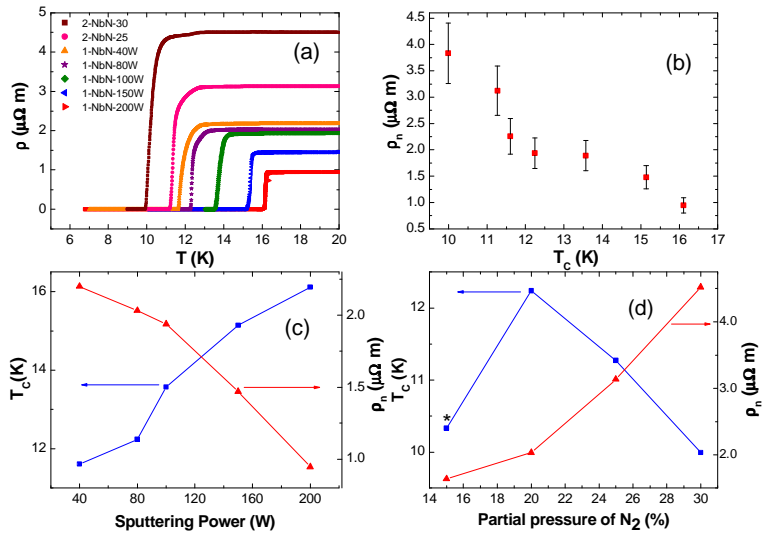

Figure 2

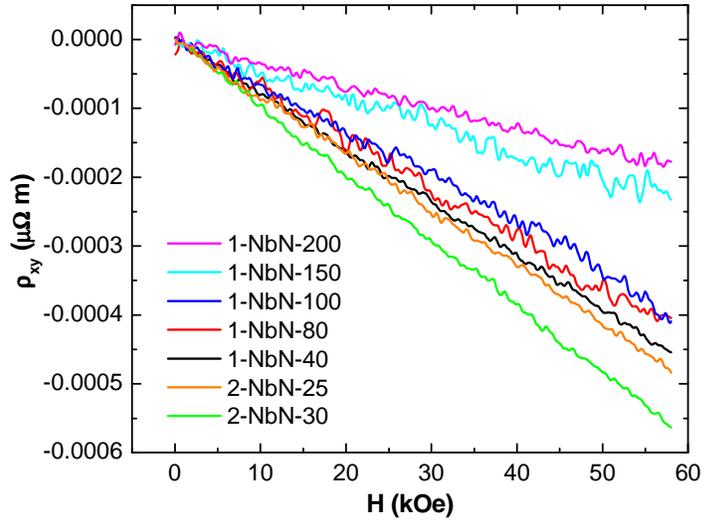

Figure 3

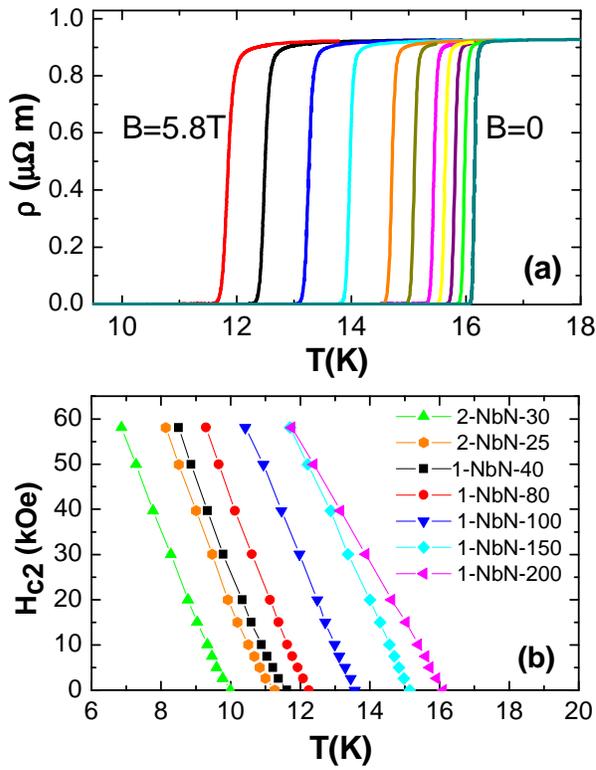

Figure 4

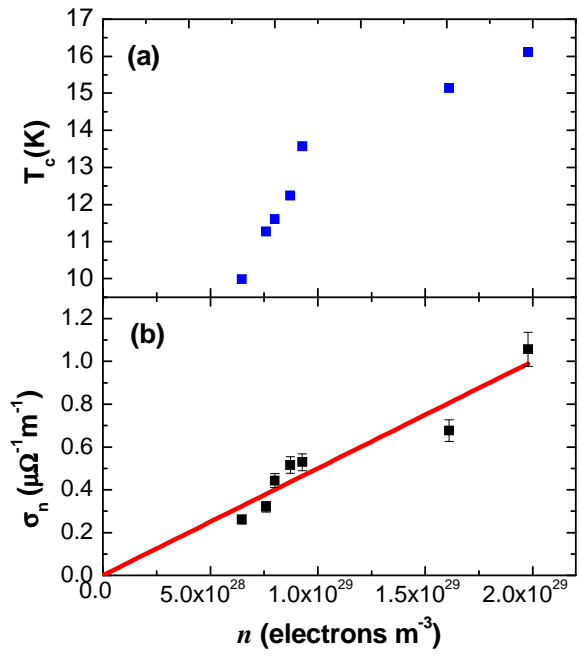

Figure 5

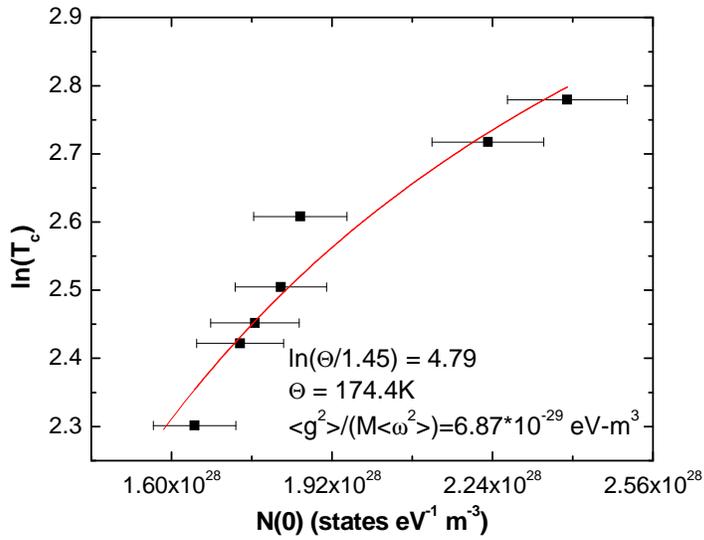

Figure 6

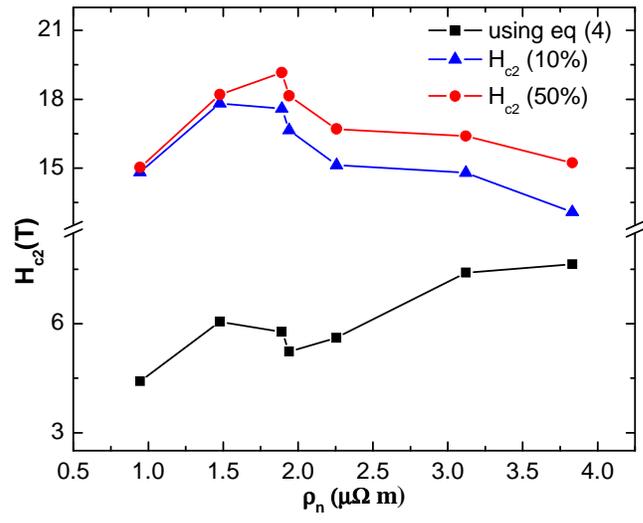

Figure 7

# References

[1] J. C. Villegier, L. Viewx-Rochaz, M. Goniche, P. Renard, M. Vabre, IEEE Trans. Magn. 21, 498 (1985); A. Kawakami, Y. Uzawa, Z. Wang, Appl. Phys. Lett. 83, 3954 (2003).

[2] S. Cherednichenko, V. Drakinskiy, J. Baubert, J.-M. Krieg, B. Voronov, G. Gol'tsman, and V. Desmaris, J. Appl. Phys. 101, 124508 (2007); P. Khosropanah, J. R. Gao, W. M. Laauwen, M. Hajenius, and T. M. Klapwijk, Appl. Phys. Lett. 91, 221111 (2007) J. J. A. Baselmans, A. Baryshev, S. F. Reker, M. Hajenius, J. R. Gao, T. M. Klapwijk, B. Voronov, and G. Gol'tsman, J. Appl. Phys. 100, 084510 (2006).

[3] Shigehito Miki, Mikio Fujiwara, Masahide Sasaki, Burm Baek, Aaron J. Miller, Robert H. Hadfield, Sae Woo Nam, and Zhen Wang, Appl. Phys. Lett. 92, 061116 (2008); M. Ejrnaes, R. Cristiano, O. Quaranta, S. Pagano, A. Gaggero, F. Mattioli, R. Leoni, B. Voronov, and G. Gol'tsman, Appl. Phys. Lett. 91, 262509 (2007); W. Sysz et al., Appl. Phys. Lett. 88, 261113 (2006).

[4] J. R. Gavaler, J. K. Hulm, M. A. Janocko and C. K. Jones, J. Vac. Sci. and Technol. 6, 177 (1968).

[5] Zhen Wang, Akira Kawakami, Yoshinori Uzawa and Bokuji Komiyama, J. Appl. Phys. 79, 7837 (1996)

[6] D. D. Bacon, A. T. English, S. Nakahara, F. G. Peters, H. Schreiber, W. R. Sinclair and R. B. Van Dover, J. Appl. Phys. 54, 6509 (1983)

[7] D. W. Capone, K. E. Gray and T. R. Kampwirth, J. Appl. Phys. 65, 258 (1989).

[8] H.-J. Hedbabny and H. Rogalia, J. Appl. Phys. 63, 2086 (1988).

[9] Y. M. Shy, L. E. Toth, and R. Somasundaram, J. Appl. Phys. 44, 5539 (1973).

[10] R. E. Treece, M. S. Osofsky, E. F. Skelton, Syed B. Qadri, J. S. Horwitz and D. B. Chrisey, Phys. Rev B 51, 9356 (1995); R. E. Treece, J. S. Horwitz, J. H. Classen and Douglas B. Chrisey, Appl. Phys. Lett. 65, 2860 (1994).

[11] K. Senapati, N. K. Pandey, Rupali Nagar and R. C. Budhani, Phys. Rev. B 74, 104514 (2006).

[12] John . Xiao and C.L. Chien, Phys. Rev. Lett 76, 1727 (1996).

[13] E.S. Sadki, Z. H. Barber, S. J. Lloyd, M. G. Blamire and A. M. Campbell, Phys. Rev. Lett 85, 4168 (2000).

[14] P. W. Anderson, J. Phys. Chem. Solids 11, 26 (1959).

[15] The notion that disorder does not affect the $T_c$ in s-wave superconductors has recently been questioned. For details see, Y. Dubi, Y. Meir and Y. Avishai, Nature 449, 876 (2007); A. Ghosal, M. Randeria and N. Trivedi, Phys. Rev. Lett. 81 3940 (1998).

[16] Very close to the limit of Mott minimum metallic conductivity the $T_c$ can decrease because of the increase in electron-electron repulsion caused by the loss of effective screening; see ref. 17.

[17] P. W. Anderson, K. A. Muttalib and T. V. Ramakrishnan, Phys. Rev. B 28, 117 (1983).

[18] Ioffe A. F. and Regel A. R., Prog. Semicond., 4 237 (1960).

[19] M. P. Mathur, D. W. Deis and J. R. Gavaler, J. Appl. Phys. 43, 3158 (1972)

[20] L. F. Mattheiss, Phys. Rev. B, 5, 315 (1969).

[21] T. H. Geballe, B. T. Matthias, J. P. Remeika, A. M. Clogston, V. B. Compton, J. P. Maita and H. J. Wiliams, Physics 2, 293 (1966).

[22] N. R. Werthhamer, E. Helfland, and P. C. Honenberg, Phys. Rev. 147, 295 (1966).

[23] For Series 2 our variation of $T_c$ with lattice parameters is similar to the one reported by Wang et al. [ref. 5].

[24] The situation here is similar to heavily doped semiconductors where the mobility becomes independent of doping concentration. See. S. C. Choo, M. S. Leong and L. S. Tan, Solid-State Electronics 24, 557 (1981).

[25] W. L. McMillan, Phys. Rev. 167, 331 (1968).

[26] K. Komenou, T. Yamashita and Y. Onodera, Phys. Lett. 28A, 335 (1968).

[27] P. Roedhammer, E. Gmelin, W. weber and J. P. Remeika, Phys. Rev. B, 15, 711 (1977).

[28] Y. Pellan, G. Dousselin, J. Pinel and Y . U. Sohn, J. Low Temp. Phys. 78, 63 (1990).

[29] C Geibel, H Rietschel, A Junod, M Pelizzone and J Muller, J. Phys. F: Met. Phys. 15, 405 (1985).

[30] We would like to note that Θ in McMillan theory arises from a phenomenological cutoff which is of the order the Debye temperature. A precise match between the two is therefore not expected.

[31] K. E. Kihlstrom, R. W. Simon and S. A. Wolf, Phys. Rev. B 32, 1843 (1985).

[32] S. Maekawa, H. Ebisawa and H. Fukuyama, Prog. Theor. Phys. Suppl. 84, 154 (1985).

[33] Since the loss of screening will affect the $N(0)$ at the Fermi level but will have no effect on the carrier density, $n$, extracted from Hall effect the loss of screening should manifest itself as a deviation from linear trend in $\sigma_n$ versus $n$ at large disorder which we do not observe in our data.

[34] The mechanism proposed by Anderson-Muttalib-Ramakrishnan (ref. 17) has however been questioned by Leavens, see C. R. Leavens, Phys. Rev. B 31, 6072 (1985).

[35] Sangita Bose, Pratap Raychaudhuri, Rajarshi Banerjee, and Pushan Ayyub, Phys. Rev. B 74, 224502 (2006).

[36] P. Marksteiner, P. Weinberger, A. Neckel, R. Zeller and P. H. Dederichs, Phys. Rev. B 33, 6709 (1986).

[37] It is interesting to note that $T_c$ decreases both for samples with excess Nb and excess N; E.K. Storms, A.L. Giorgi and E.G. Szklarz, Journal of Physics and Chemistry of Solids 36, 689 (1975).

[38] The situation here is possibly similar to the Ar ion irradiated NbN samples where the decrease in $T_c$ after irradiation was attributed to the reduction in density of states caused by Nb or N vacancies; J. Y. Juang, D. A. Rudman, J. Talvacchio, R. B. van Dover, Phys. Rev B 38, 2354 (1988).

[39] T. Ohashi, H. Kitano, A. Maeda, H. Akaike and A Fujimaki, Phys. Rev. B. 73, 174522 (2006).